# Specific Heat Investigation for Line Nodes in Heavily Overdoped $Ba_{1-x}K_xFe_2As_2$

J. S. Kim[1], G. R. Stewart[1], Yong Liu[2], and Thomas A. Lograsso[2,3]

[1]Department of Physics, University of Florida, Gainesville, FL 32611;  [2]Division of Materials Sciences and Engineering, Ames Laboratory, US DOE, Ames, Iowa 50011, USA; [3]Department of Materials Science and Engineering, Iowa State University, Ames, Iowa 50011, USA

**Abstract:**  Previous research has found that the pairing symmetry in the iron-based superconductor $Ba_{1-x}K_xFe_2As_2$ changes from nodeless s-wave near optimally doped, x≈0.4-0.55 and $T_c$>30 K, to nodal (either d-wave or s-wave) at the pure endpoint, x=1 and $T_c$<4 K.  Intense theoretical interest has been focused on this possibility of changing pairing symmetry, where in the transition region both order parameters would be present and time reversal symmetry would be broken.  Here we report specific heat measurements in zero and applied magnetic fields down to 0.4 K of three individual single crystals, free of low temperature magnetic anomalies, of heavily overdoped $Ba_{1-x}K_xFe_2As_2$, x= 0.91, 0.88, and 0.81.  The values for $T_c^{mid}$ are 5.6, 7.2 and 13 K and for $H_{c2}$ ≈ 4.5, 6, and 20 T respectively.  The data can be analyzed in a two gap scenario, $\Delta_2/\Delta_1 \approx 4$, with the magnetic field dependence of γ (=C/T as T→0) showing an anisotropic 'S-shaped' behavior vs H, with the suppression of the lower gap by 1 T and $\gamma \approx H^{1/2}$ overall.  Although such a non-linear γ vs H is consistent with deep minima or nodes in the gap structure, it is not clear evidence for one, or both, of the gaps being nodal in these overdoped samples.  Thus, following the established theoretical analysis of the specific heat of d-wave cuprate superconductors containing line nodes, we present the specific heat normalized by $H^{1/2}$ plotted vs $T/H^{1/2}$ of these heavily overdoped $Ba_{1-x}K_xFe_2As_2$ samples which – thanks to the absence of magnetic impurities in our sample - convincingly shows the expected scaling for line node behavior for the larger gap for all three compositions.  There is however no clear observation of the nodal behavior $C \propto \alpha T^2$ in zero field at low temperatures, with α ≤ 2 mJ/molK$^3$ being consistent with the data.  This, together with the scaling, leaves open the possibility of extreme anisotropy in a nodeless larger gap, $\Delta_2$, such that the scaling works for fields above 0.25 – 0.5 T (0.2 – 0.4 K in temperature units), where this an estimate for the size of the deep minima in the $\Delta_2$ ~ 20-25 K gap.   Therefore, the location of the change from nodeless→nodal gaps between optimally doped and heavily overdoped $Ba_{1-x}K_xFe_2As_2$ based on the present work may be closer to the $KFe_2As_2$ endpoint than x=0.91.

I. Introduction

Although Rotter, Tegel and Johrendt discovered[1] superconductivity in the second iron-based superconductor structure, 122 $BaFe_2As_2$ doped with K on the Ba site, only a short time after Hosono and coworkers' seminal discovery[2] in fluorine doped 1111 LaFeAsO, the properties of samples in the $Ba_{1-x}K_xFe_2As_2$ phase diagram continue to be of interest seven years later. Surprisingly, the focus continues to be primarily near the $KFe_2As_2$ endpoint, where $T_c$ is less than 4 K, the specific heat γ is[3] of order 100 mJ/molK$^2$, and the discontinuity in the specific heat at $T_c$, ΔC, is anomalously large.[3-4] It is generally accepted that the pairing symmetry near optimally doped $Ba_{0.6}K_{0.4}Fe_2As_2$, and up to x=0.55, is[5] s-wave, with no accidental nodes, and that $KFe_2As_2$ is nodal (either d-[6-7] or s-wave[8-9]). Thus, an important focus issue is how does the nodal behavior evolve between x ≈ 0.55 and 1.0? Theoretical studies[10-12] have discussed how this transition, or 'mixed' region, where the pairing symmetry might change from s± to $d_{x2-y2}$-pairing, could be in an s+id state and exhibit, among other interesting phenomena, breaking of time reversal symmetry. Even if the transition region is just between an s± state with electron and hole pockets (optimally doped) and the x=1, purely hole pocket s± state, it has been argued[13-15] that this s+is transition regime could also exhibit time reversal symmetry breaking.

This work presents an analysis of low temperature specific heat, C, data in fields up to 12 T to look for the presence of nodes in single crystals of $Ba_{1-x}K_xFe_2As_2$, with $T_c^{mid}$ values=5.6, 7.2 and 13 K (called 'samples #1, #2, and #3' hereafter). As will be discussed, this analysis follows a number of both theoretical and experimental works that developed this technique in verifying the existence of d-wave line nodes in the cuprate superconductors. If data plotted as $C/H^{1/2}$ at temperatures much less than $T_c$ scale onto one curve when plotted vs $T/H^{1/2}$, this implies line node behavior.

A number of iron based superconductors contain magnetic impurities which then produce[16] Schottky anomaly magnetic responses in zero and applied fields at low temperatures which make such analysis for the presence of nodes either difficult or impossible. Thus, similar analysis in YBCO crystals, the canonical nodal superconductor, was difficult for years until finally better samples[17] could be prepared, and even then one of the important parameters still has a large uncertainty. Fortunately – unusually for K-doped $BaFe_2As_2$[18] - the crystals for this work contains minimal or no such impurities, and the analysis can proceed straightforwardly.

## II. Experimental

Overdoped $Ba_{1-x}K_xFe_2As_2$ ($x$=0.80-0.95) single crystals were grown by the KAs flux method[19-20]. The starting materials of Ba and K lump, and Fe and As powder were weighed at a ratio of Ba:K:Fe:As=$y$:5:2:6 ($y$ =0.1 and 0.2). The chemicals were loaded into an alumina crucible, and then sealed in a tantalum tube by arc welding. The tantalum tube was sealed in a quartz ampoule to prevent the tantalum tube from oxidizing in the furnace. Thin plate-like single crystals with up to 1 centimeter in size were obtained utilizing a cooling rate of 3 K/h from 1323 K to 1173 K and 1 K/h from 1173 K to 1023 K. Single crystals were carefully cleaved along the *ab* plane. Superconducting transitions were measured with a Vibrating Sample Magnetometer (VSM) for the Physical Property Measurement System by Quantum Design. The actual composition of the three chosen crystals was determined with a wavelength dispersive x-ray spectroscopy (WDS) detector in an SEM. Three different compositions of $Ba_{1-x}K_xFe_2As_2$, $x$= 0.91, 0.88, and 0.81 with

$T_c^{mid}$ = 5.6, 7.2 and 13 K respectively, (labeled samples 1, 2 and 3 herein) were chosen for this study.

Specific heat was measured according to established methods.[21] In order to insure a minimal error bar (±3%), three different masses (4.54 mg, 7.14 mg, and 29.01 mg) of an ultra-high purity Au standard obtained from NIST were measured first.

## II. Results

The specific heat of the three samples is shown in Fig. 1. As can be seen, the impurity upturn in C/T often seen below 1 K in the $Ba_{1-x}K_xFe_2As_2$ system is vanishingly small (see also Figs. 2-4 below) in all three single crystal samples. Secondly, all three samples show a strong 'shoulder' feature in C/T at low temperatures indicative of a second superconducting energy gap as seen in, e. g., $MgB_2$, and as discussed below.

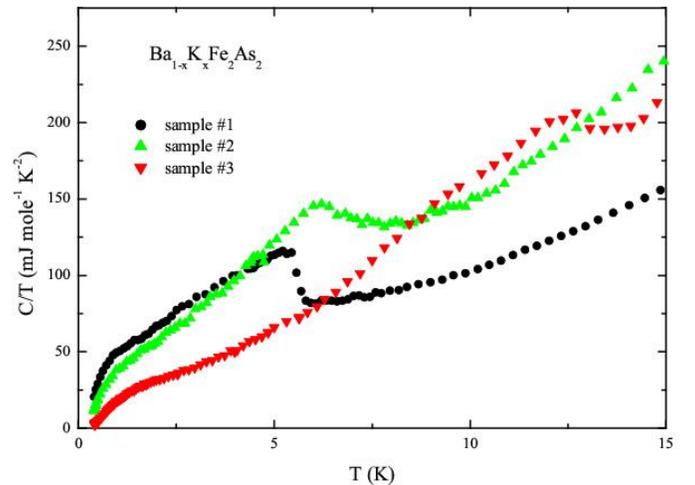

Fig. 1 (color online) Specific heat, C, of three samples of $Ba_{1-x}K_xFe_2As_2$ divided by temperature T vs T. The error bar (not shown) of the data is ±3% up to 10 K. Due to the small masses of the crystals (<5 mg), the addenda contribution to C exceeds 50% above 10 K and these data are less accurate. For the scaling analysis presented below, only data at 2.1 K and below are used. In general, the behavior of sample 2 will be intermediate between the behaviors of samples 1 and 3 because of its intermediate $T_c$; thus, its properties will be mentioned but, for brevity, not graphically presented past its appearance here in this figure.

The specific heat, C, corrected for the low temperature hyperfine field splitting of nuclear levels (important only below ~1 K in fields to 12 T) is denoted ΔC. In order to eliminate the lattice contribution, which is not field dependent, ΔC/T as a function of field vs temperature, T, for sample #1 is plotted in Fig. 2a with ΔC/T (4.5 T) subtracted. Since H=4.5 T is rather close to $H_{c2}$ for this sample, this ΔC/T (4.5 T) will have essentially no contribution from the vortices. This subtraction works well to eliminate $C_{lattice}$ due to the essentially total absence of a Schottky anomaly at low temperatures in the sample which, if present, would be field dependent and preclude using the C/T (4.5 T) data as a lattice subtraction.

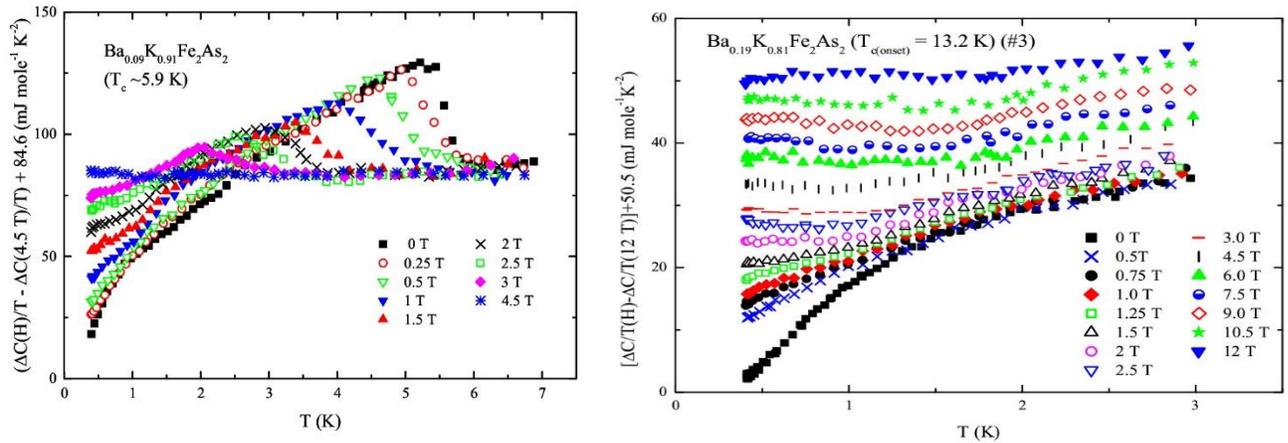

Fig. 2 (color online) a.) Specific heat of a 4.77 mg single crystal of $Ba_{0.09}K_{0.91}Fe_2As_2$ corrected for the low temperature hyperfine field contribution, ΔC, divided by temperature as a function of field, with the 4.5 T data subtracted to eliminate the lattice specific heat, is plotted vs temperature. In order to plot against a positive vertical axis, these difference data are shifted upwards by a constant equal to the T→0 value (84.6 mJ/molK$^2$) of the subtracted ΔC/T (4.5 T) curve. Thus, this is a plot of the electronic superconducting specific heat, $C_{el}$. H=0.125 T not shown for clarity. b.) Low temperature specific heat of a 4.68 mg single crystal (sample #3) of $Ba_{0.19}K_{0.81}Fe_2As_2$ corrected for the low temperature hyperfine field contribution, ΔC, divided by temperature as a function of field (with 12 T data subtracted to eliminate the lattice specific heat and ΔC/T (12 T, T→0) = 50.5 mJ/molK$^2$ added to show the electronic superconducting specific heat, $C_{el}$) is plotted vs temperature. This procedure for sample #3 does not involve the full normal state γ since $H_{c2}$~20 T. However, the 12 T data suffice to subtract away the lattice contribution (which is not field dependent and is therefore exactly canceled by this procedure) without reducing severely the vortex contribution in the difference specific heat.

The same procedure is also followed for samples 2 (H=6 T) and 3 (H=12 T), with the constants equal to 79.6 and 50.5 mJ/molK$^2$ respectively, with the result for sample #3 shown in Fig. 2b. Thus, our discussion of the specific heat of $Ba_{1-x}K_xFe_2As_2$, x=0.91, 0.88, and 0.81 (samples 1-3), below will depend on the *difference* specific heat.

Although flux pinning has in other samples, e. g. in the electron-doped cuprate superconductor PCCO, $T_c$=22 K caused[22-23] confusing differences in field cooled vs zero field cooled specific heat measurements, in the present work direct measurements of both field and zero field cooled data have shown no measureable differences down to 0.4 K.

Before we discuss the extrapolation of C/T to T=0, let us discuss the zero field data in Figs. 1 and 2. First, note that there is no upturn at low temperatures in C/T, i. e. no indication of impurity phases.[24] Second, the round shoulder in C/T at around 1 K (samples #1 and #2) and 1.5 K (sample #3) clearly is indicative of a second gap as seen in, e. g., $MgB_2$.[25] Fig. 3 below shows a two gap fit to the zero field data for sample #1 as an example, with the result that the ratio of the larger (black line) gap to the smaller gap (blue line) fit is $\Delta_2/\Delta_1 \approx 3.9$. Although the fit to the data is improved by the addition (not shown) of a term $\alpha T$ to C/T with $\alpha$=1 mJ/molK$^3$ (the magnitude of the $\alpha$ coefficient chosen comes from discussion below), this is in no way definitive.

Similar fits (not shown) for samples #2 and 3 result in $\Delta_2/\Delta_1 \approx 3.9$ and 4.4 respectively. From Figs. 1 and 2, the shoulder in C/T for sample #1 below ~ 1 K, caused by the lower gap, moves up gradually in temperature with the increasing $T_c$ onsets in the progression of the compositions from sample 1 to 3. Thus, the lower gap $\Delta_1/k_B$ for sample 1 is slightly less than 0.5 that for sample 3, corresponding visually with the relative temperatures of the shoulders in the low temperature specific heat in Figs. 1 and 2. This higher $\Delta_1$ for sample #3 will be important in the discussion of scaling below. The presence of this second, smaller gap in the samples complicates the investigation of possible nodal behavior. The original specific heat evidence[26] for d-wave pairing in YBCO, i. e. that $\gamma \propto H^{1/2}$, can be mimicked[27] in fully gapped multiple gap systems, thus removing measurement of γ as a function of field as a method for conclusively indicating nodal behavior. However, as will be discussed below, the application of rather small magnetic fields can suppress the lower gap, allowing scaling analysis of the specific heat from the larger gap for nodal (or at least very deep minima) behavior.

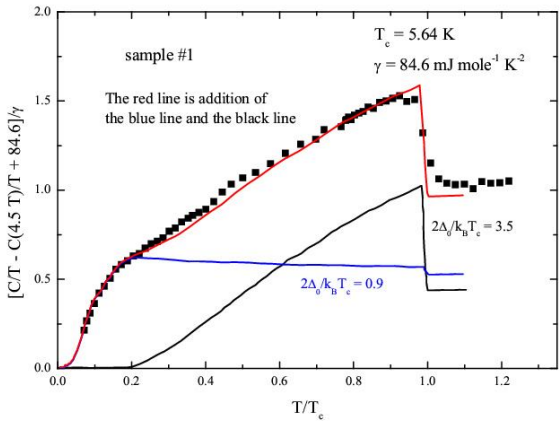

Fig. 3 (color online) Two gap fit to the zero field specific heat of $Ba_{0.09}K_{0.91}Fe_2As_2$. The shoulder in the zero field specific heat data in Fig. 1 around 1 K is due to a rather small gap, $\Delta_1/kT_c = 0.45$. Since the fit (black line) to the larger gap results in essentially zero specific heat below $T/T_c=0.2$, the red (sum of both fits) and the blue (fit to the smaller gap) lines coincide (cannot be distinguished) below $T/T_c=0.2$. There is no correction to C(0) for the hyperfine contribution $\propto H^2/T^2$.

A. <u>Extrapolation of ΔC/T to T=0 from 0.4 K :</u>  Before we discuss the γ vs H analysis, it is important to discuss how the extrapolation of C/T (T→0) to determine γ is done. Although the extrapolation of the low temperature specific heat data in each field (see Fig. 2) should be straightforward due to the short interval from 0.4 K to 0 K, in order to minimize the possibility that the method chosen to extrapolate introduces a bias (for example, making a linear fit through C/T vs T data could be argued to favor a line node interpretation, since nodal s- or d-wave behavior should create a (small) $C \propto T^2$ term), we have adopted the following complementary schemes. First, at each field (see Fig. 4

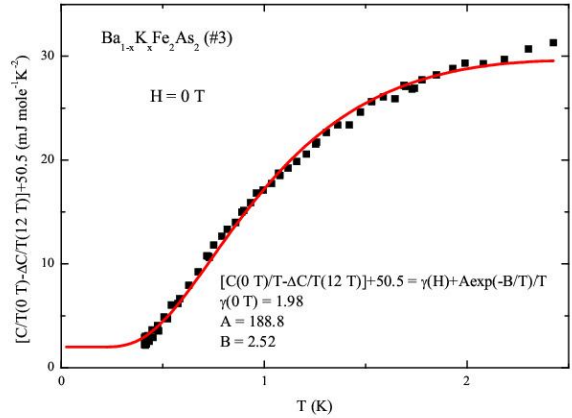

Fig. 4 (color online): BCS exponential fit to the zero field data of sample #3 at low temperatures to determine γ. Our observed residual linear term is less than 5% of $\gamma_{normal}$. This value of $\gamma_{residual}$ is less than the 15% of $\gamma_{normal}$ value in the clean limit YBCO sample of Wang et al.[17]

for an example at 0 T), we fit the low (≤1.5 K) temperature data from Fig. 1 to the form γ(H) + [aexp(-b/T)]/T to find γ as a function of field for each field (as done by Bouquet et al.[25] for $MgB_2$). The parameters a and b are independently chosen at each field to give the best fit. As clear from Fig. 4, this fit is a good representation of the low temperature zero field data. As can be seen in Fig. 1 and also here in Fig. 4, the low temperature C/T data for H=0 – due to the opening of the lower gap, $\Delta_1$, - has a strong exponential temperature dependence which makes it difficult to directly fit C/T to an αT term. As can also be seen in Fig. 2, starting already at a field of 0.5 T this is no longer true.

Of course, using a fitting form assuming a full gap, like the one shown in Fig. 4, could also be claimed to introduce a bias in our determination of γ. Thus, our second scheme for obtaining γ vs H is simply not to do an extrapolation, and just plot C/T (0.4K) vs H, since the T=0.4 K specific heat data may be considered as representative of γ.

Note that all three compositions of $Ba_{1-x}K_xFe_2As_2$ (as is also observed[28] in, e. g., Co-doped $BaFe_2As_2$) have finite γ's (C/T as T→0) in the superconducting state (~ 2

mJ/molK$^2$ for sample #3, Fig. 4). Although the cause of this ubiquitous finite '$\gamma_{residual}$' is still under discussion, its presence prevents any discussion of assigning a small finite $\gamma$ as indicative of the presence of line nodes.

**B. $\gamma$ vs H:** Fig. 5 shows the resultant $\gamma$ vs H graph for sample #3, using the $\gamma$(H) + [aexp(-b/T)]/T form for one fit (blue curve), and presenting C/T (0.4K) vs H, with no extrapolation, (black curve) for comparison. As can be seen, both sets of data are quite similar and show clear indication of a two gap scenario, with the lowest gap being fully suppressed in the range H=1 – 1.5 T. Whether the higher field (> 1 T) sub-linear behavior of $\gamma$ with H can be described as having deeper import than just being consistent with deep minima in the gap function, or possibly even nodes[29], cannot be determined without other measurements or analysis. This is because in a multiband superconductor such sub-linear $\gamma$ vs H can also be from fully gapped s-wave behavior[27] such as seen in Nb[30] or MgB$_2$[25]. Thus, we procede to a scaling analysis of the specific heat data in the next section in order to better determine if there are nodes present in Ba$_{1-x}$K$_x$Fe$_2$As$_2$.

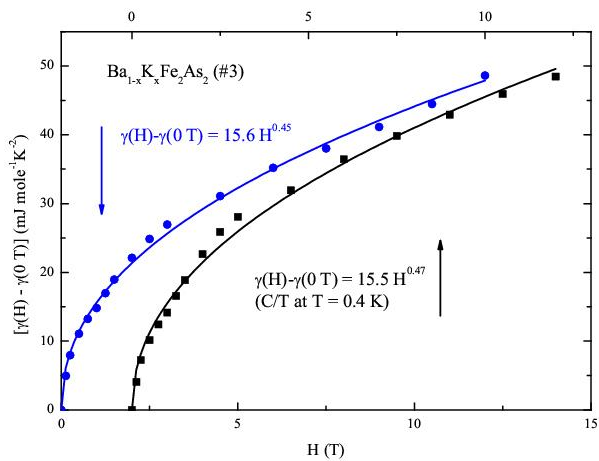

**Fig. 5 (color online)** $\gamma$ extrapolated to T=0 using a gapped aexp(-b/T) fit form to the low temperature specific heat data for sample 3 below 1.5 K, where a and b are independently determined at each field (blue curve) as well as C/T(H,0.4 K) (black curve) vs H. Note the suppression of the lower gap at about 1.5 T, resulting in the slight 'S' shape of $\gamma$ vs H, indicative of two gaps. The horizontal (H) axis for the C/T(H, 0.4 K) data (upper axis) has been shifted for clarity.

**A. Scaling of $[\Delta C(H) - C(0)]/T*H^{1/2}$ vs $T/H^{0.5}$:** Volovik[31] was the first to point out that in finite magnetic fields above $H_{c1}$ (i. e. in the mixed state) in a superconductor with line nodes (which in iron based superconductors can come either from d-wave pairing where the nodes are symmetry driven, or from s-wave pairing where the nodes are accidental), there was a Doppler shift of the quasiparticle excitation spectrum. This shift in the neighborhood of the nodes is of the order of the superconducting energy gap, and therefore strongly alters the density of states at the Fermi energy, which can be probed by measuring the specific heat. There are two regimes to consider: zero field/very low temperature gives a term in the electronic specific heat, $C_{el,} = \alpha T^2$ (where reported values for $\alpha$ in nodal d-wave YBCO derived from this kind of scaling analysis are[17,32] between 0.044 and 0.21 mJ/molK$^3$, i. e. somewhat uncertain and very small) and at zero temperature/very low field $C_{el} = A_c T H^{1/2}$. (The low field, $C_{el}/T \propto H^{1/2}$ behavior derived by Volovik gets its field dependence from the field dependence of the inter-vortex spacing and is the same for line nodes from d-wave or from s± pairing symmetry.) These two limiting formulae, discussed in various theoretical works,[31,33-36] are equal at a crossover temperature. This crossover temperature $T_{cross}$ is a function of H, with $A_c/\alpha = T_{cross}(H)/H^{1/2}$, which is equal to a numerical constant[17,34] times $av_F$, where 'a' is a constant of order 1 and $v_F$ is the Fermi velocity. Thus, $T_{cross}(H) \propto H^{1/2}$. (Analysis based on the d-wave theories[31,33-36] below will give us, within constants dependent on d vs s± symmetry, an idea of $T_{cross}(H)$ for $Ba_{1-x}K_xFe_2As_2$.)

Analytical theoretical details in the crossover regime are as yet unknown, but $[C_{el}/\gamma_{normal}T]*(H_{c2}/H)^{1/2}$ scales[17,34] as a function $F(\{T/T_{cross}(H)\})$. As well, Kuebert and

Hirschfeld have determined[33] the scaling function F numerically for the low energy clean limit for d-wave superconductors. Although the correct analytic interpolating function $F(x)$, $x=T/T_{cross}(H)$, is under some discussion[36], if (in the nomenclature of the present work) the data plotted as $[\Delta C(H) - C(0)]/TH^{1/2}$ vs $T/H^{1/2}$ collapse onto a single curve, this scaling is then consistent with the existence of line nodes – whether from d-wave or from s± symmetry - as presented in theories of Simon and Lee[34], Kuebert and Hirschfeld[33], and Vekhter et al[35-36] for d-wave superconductors. It is important to sample more than just the crossover regime to have a proper (wide in parameter space) check of the scaling, i. e. to temperatures at least below $T_c/10$ and in fields at least below $H_{c2}/10$. Since $T_c^{mid}$ in the present samples of $Ba_{1-x}K_xFe_2As_2$ is 5.6, 7.2, and 13.2 K and $H_{c2}$ is between 4.5 and approximately 20 T, having temperatures below 0.6 K and fields ≤ 0.5 T allows us to reach far below the crossover regime in all three samples. Thus, in order to check for this scaling, we plot $[\Delta C(H) - C(0)]/T*H^{1/2}$ vs $T/H^{0.5}$ for $Ba_{0.09}K_{0.91}Fe_2As_2$ in Fig. 6a and for $Ba_{0.19}K_{0.81}Fe_2As_2$ in Fig. 6b at various low fixed temperatures and fields of 0.125 - 4.5 T for sample 1 and 0.5 – 12 T for sample 3. The scaling for sample 2 (not shown) has points for

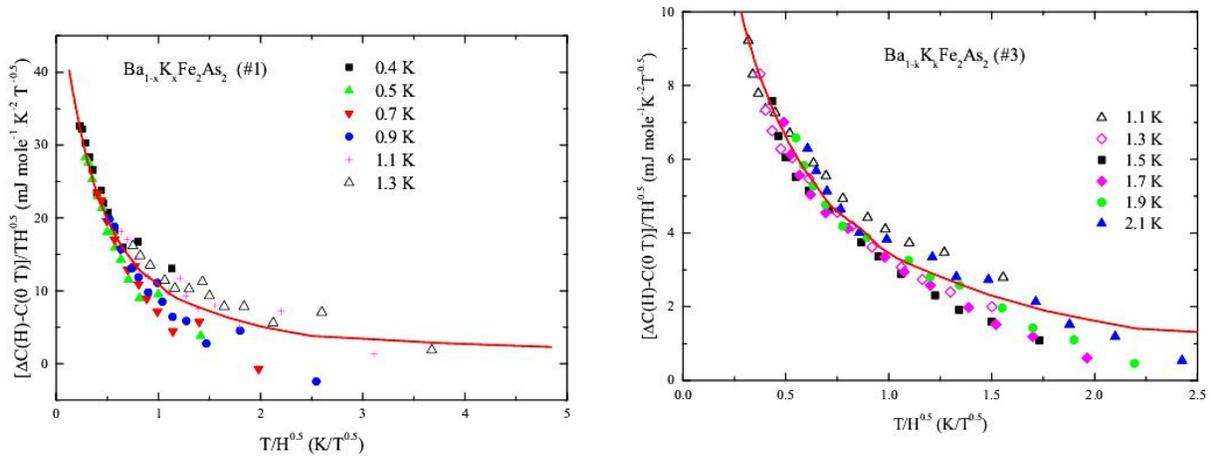

**Fig. 6 (color online)** Data plotted as $[\Delta C(H) - C(0)]/TH^{1/2}$ vs $T/H^{1/2}$ collapse onto a single curve for samples #1, #2 (not shown), and #3, (T=0.4, 0.5, 0.7, 0.9, and 1.1 K for sample 1, 0.6 – 1.2 K for sample 2, and 1.1-2.1 K for sample 3) indicative of line nodes in the larger of the two gaps in this composition range ($0.81 \leq x \leq 0.91$) of $Ba_{1-x}K_xFe_2As_2$. The temperatures are chosen so that the data scale onto one curve for each sample; as the smaller gap grows in temperature with the decreasing amount of potassium and dominates the lowest temperature data, these lowest temperature data therefore do not scale with the scaling data of the larger gap. Note that, unlike in work on cuprate superconductors, e. g. YBCO[17] or earlier work on LSCO[37] ($La_{2-x}Sr_xCuO_4$), these data did not need to be corrected for any Schottky contribution due to the quality of the samples in the present work. The solid red line in each graph is the numerically derived interpolation function of Kuebert and Hirschfeld[33], scaled by a constant for both the vertical and horizontal axes, as discussed in the text.

fields between 0.125 and 6 T. These data scale onto one curve as shown in Figs. 6a and 6b for samples 1 and 3. The plotting of this difference specific heat follows the treatment of Wang et al.[17] for analyzing d-wave line node behavior in the specific heat of YBCO and avoids any fit to the background contributions (except for the straightforward subtraction of the hyperfine field contribution, $C \sim H^2/T^2$, to the lowest temperature, T<1 K, data). As mentioned above, the following discussion only relies on differences in the specific heat.

As apparent in Fig. 6a, the difference $[\Delta C(H) - C(0)]/T$ in $Ba_{0.09}K_{0.91}Fe_2As_2$, normalized by $H^{1/2}$, scales rather well with $T/H^{1/2}$ (as predicted[33-36] for a superconductor with line nodes) and with the numerical fit of Kuebert and Hirschfeld[33] for temperatures almost up to 0.2 $T_c$, although the 1.1 K points start to deviate from the common curve at the lowest fields. This deviation increases for temperatures above 1.1 K. For sample 3 (Fig. 6b), with the exception of the lowest temperature data (0.4 - 0.9 K) below the lower gap, $\Delta_1$, all of the data scale onto one curve up to 2.1 K, ~ 0.16 $T_c$. The lowest temperature regime for sample #3 still has a contribution from the lower gap evident in the low (< 1 K) temperature specific heat in zero field shown in Fig. 2b. This contribution in the temperature range 0.4-0.9 K is much stronger than in sample 1, which as discussed above with Fig. 3 has a lower gap $\Delta_1$ only about half that of sample 3. As a working hypothesis, the effect of field on the vortices associated with this lower gap transition does not scale with the data involving the larger gap. Presumably this is simply due to the much different energetics involved, since even these low fields and temperatures are much larger fractions of the characteristic values for the lower gap transition than for the larger gap portion of the Fermi surface (e. g. 0.5 - 1 T is approximately the critical field for the small gap). As seen in Fig. 6b and in the lower temperature data in Fig. 2, the effect of even just 0.5 T below 1 K on the specific heat of sample 3 from the lower transition is enormous compared to the effect of field at higher temperatures. By 1.1 K (open black triangles), the difference data for sample 3 are already a perfect match with the other data.

This collapse of the data onto a single scaling curve therefore supports the existence of line nodes in the larger gap sheet of the Fermi surface in $Ba_{1-x}K_xFe_2As_2$, x>0.8. As to whether these nodes are symmetry imposed (d-wave pairing symmetry) or accidental (as in an s-wave superconductor) cannot be distinguished by this analysis. It is interesting to note that a similar effort[38] to scale C(H, T) data for optimally doped $Ba_{0.6}K_{0.4}Fe_2As_2$ to follow the nodal scaling predictions of refs. 33-36 was unsuccessful, implying fully gapped behavior as is believed[5] from other measurements.

Fig. 7, a plot of the difference specific heat at various temperatures from Fig. 6 vs $H^{1/2}$, shows how the parameters $A_c$ and $\alpha$ can be extracted from the data. (These parameters are necessary to calculate the crossover temperature, $T_{cross}= (A_c/\alpha)*H^{1/2}$.) Considering Fig. 7b for sample 3 as an example, at higher fields (e. g. ≥3 T for 1.1 K, and ≥5 T for 2.1 K), the data in Fig. 6 lie on parallel straight lines. This is the region where x (=T/$T_{cross}$(H)) $\propto H^{-1/2}$, is much less than 1 and[17] the difference specific heat divided by

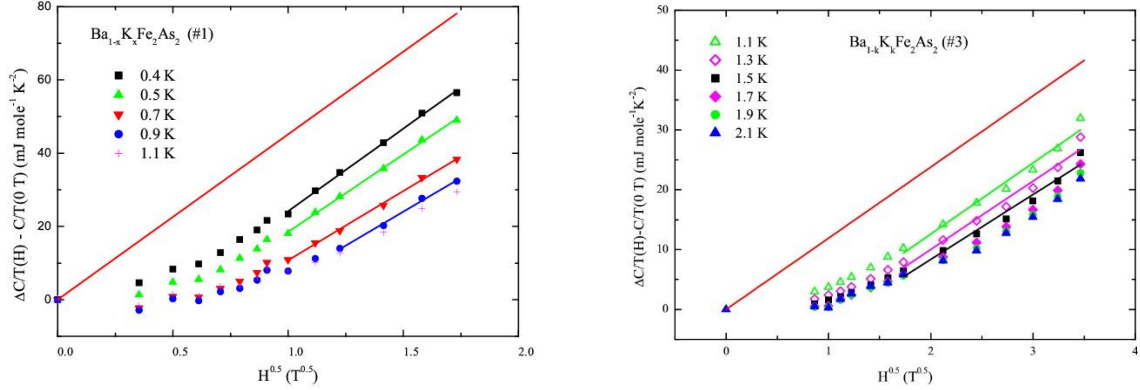

**Fig. 7 (color online)** $[\Delta C(H) - C(0)]/T$ is plotted vs $H^{1/2}$. As the data go to higher T the region where $x=T/T_{cross}(H) \sim H^{-1/2}$ is much less than 1, i. e. where the plotted data lie on a straight line, obviously is restricted to the highest fields. The red line through 0,0 is a guide to the eye.

temperature varies as $A_c H^{1/2}$. Thus, the slope of the higher field difference specific heat divided by temperature data in Fig. 7a/b for sample 1/3 gives $A_c \approx 45/13$ mJ/molK$^2$T$^{1/2}$.

As discussed above, $C \to \alpha T^2$ in the zero field, low temperature limit, so that the lowest field data in Fig. 7 can be extrapolated for each temperature to H=0 to give values[39] for $\alpha$. For example for sample #3, these extrapolated values for $\alpha$ vary between 0.5 and 4 mJ/molK$^3$. From Wang et al.[17], for d-wave pairing we have

$$A_c/\alpha = T_{cross}(H)/H^{1/2} = [4\pi\hbar/(27\zeta(3)\Phi_0^{1/2}k_B)]av_F$$

i.e. the ratio of these two parameters determined from Fig. 7 depends only on constants (which will be different for the present work's overdoped BaFe$_2$As$_2$ if the line nodes are due to accidental nodes from s± pairing) and the Fermi velocity. Using the constants given in the above equation for d-wave pairing, $av_F$ (with 'a' a constant of order 1) would be, e. g. for sample #3 using $\alpha=2$ mJ/molK$^3$, $1 \; 10^7$ cm/s.

Fermi velocities reported[40] for various bands along the Γ-M direction in Ba$_{0.6}$K$_{0.4}$Fe$_2$As$_2$ range between 3.1 and 7.5 10$^6$ cm/s and Reid et al.[6] cite an average Fermi velocity in KFe$_2$As$_2$ of 4 10$^6$ cm/s. Thus, without knowing the constant 'a' and the correct constants in the above equation in the case that our Ba$_{1-x}$K$_x$Fe$_2$As$_2$ has line nodes due to s± symmetry, our calculated v$_F$ is roughly consistent with measured values.

Using the values calculated from Fig. 7 for A$_c$ and α, T$_{cross}$/H$^{1/2}$ (=A$_c$/α) ≈6-10 K/T$^{1/2}$ for samples 1, 2 and 3. Thus, the temperatures and fields for the scaling data in Fig. 6 probe the region for x (=T/T$_{cross}$, where T$_{cross}$ = 6-10 K/T$^{1/2}$ * H$^{1/2}$), up to about x=0.5. As discussed above, as the temperature (∝ x) increases to a larger fraction of T$_c$ in the scaling plots of Fig. 6, the points begin to diverge from the common scaling curve. As a comparison, Wang et al.[17] in their specific heat scaling study of YBCO reported data up to about 4% of T$_c$, and fields down to 0.16 T, allowing them, with a similar A$_c$/α, to reach x=1.6. In the case of the present work, fields smaller than 0.5 T for sample 3, which would have reached the crossover regime x=1, begin to introduce the lower energy gap with its different scaling into the data. Another instance in the cuprates worthy of comparison is the work[41] (somewhat more recent than that in ref. 37) on high quality (no Schottky upturn in C/T) LSCO samples. In that work, the underdoped samples do not show good nodal scaling (although optimally doped does), and the explanation[41] may be a competing order (e. g. antiferromagnetism) whose field dependence is different than that in the nodal quasiparticle scaling of Simon and Lee[34]. This is similar to the competition (restricted to low fields and temperatures) in the present work from the field dependence of C(H, T) from the lower energy gap.

**Conclusions**

**Specific heat data on three clean single crystals of $Ba_{1-x}K_xFe_2As_2$, x=0.91, 0.88 and 0.81 ($T_c^{onset}$ = 5.9, 7.2, and 13.2 K) show two gap behavior with a ratio between the gaps of approximately a factor of four. Except for the very low temperature/low field regime (where the lower gap dominates the specific heat response to field), scaling of the field and temperature dependence of the specific heat shows conclusive evidence for nodal behavior, or deep minima down to the scale of ~ 0.4 K, in the larger gap. Thus, the nodeless behavior found up to $Ba_{1-x}K_xFe_2As_2$, x=0.55, changes at least over to deep minima behavior by x=0.81. Whether a measureable $C \sim \alpha T^2$ will appear as x→1 in $Ba_{1-x}K_xFe_2As_2$, with the inherent masking of the low energy temperature/field scales by the smaller gap intrinsic to this compound, is a subject for further investigation – presumably to begin with in pure $KFe_2As_2$. This is the first successful application of the Volovik et al. theory of the influence of line nodes on the specific heat of a superconductor where the pairing symmetry may be[8-9] s-wave.**

Acknowledgements: Very helpful discussions with Peter Hirschfeld and Ilya Vekhter are gratefully acknowledged. Work at Florida performed under the auspices of the U. S. Department of Energy, Basic Energy Sciences, contract no. DE-FG02-86ER45268. Work at Ames Laboratory was supported by the U.S. Department of Energy, Office of Basic Energy Sciences, Materials Science and Engineering Division. Ames Laboratory is operated for the U.S. Department of Energy by Iowa State University under Contract No. DE-AC02-07CH11358.